\documentclass[twocolumn,aps,prl,superscriptaddress,floatfix]{revtex4}

\usepackage{amsfonts}
\usepackage{amsmath}
\usepackage{amssymb}
\usepackage{graphicx}
\usepackage{nicefrac}
\usepackage{tikz}

\newcommand{\dd}{\, \mathrm d}
\newcommand{\ii}{\, \mathrm i}
\newcommand{\pdag}{{\phantom\dagger}}
\newcommand{\hc}{\text{h.c.}}

\newcommand{\charge}{\text{c}}
\newcommand{\spin}{\text{s}}

\newcommand{\uni}{\text{u}}
\newcommand{\osc}{\text{o}}

\newcommand{\kF}{k_{\text F}}

\newcommand{\ket}[1]{\left| #1 \right. \rangle}
\newcommand{\bra}[1]{\langle \left. #1 \right|}

\renewcommand\Im{\operatorname{Im}}

\begin{document}

\title{Low-energy local density of states of the 1D Hubbard model}
\author{Stefan A.~S\"{o}ffing}
\affiliation{Dept.~of Physics and Research Center OPTIMAS, Univ.~Kaiserslautern, D-67663 Kaiserslautern, Germany}
\affiliation{MAINZ Graduate School of Excellence}

\author{Imke Schneider}
\affiliation{Institut f\"ur Theoretische Physik, Univ. Dresden, D-01062 Dresden, Germany}

\author{Sebastian Eggert}
\affiliation{Dept.~of Physics and Research Center OPTIMAS, Univ.~Kaiserslautern, D-67663 Kaiserslautern, Germany}
\affiliation{MAINZ Graduate School of Excellence}


\begin{abstract}
We examine the {local} density of states (DOS) at low energies 
numerically and analytically for the Hubbard model in one dimension.
The eigenstates represent
separate spin and charge excitations with a remarkably rich 
structure of the local DOS 
in space and energy. 
The results predict signatures of strongly correlated excitations 
in the tunneling probability along finite quantum wires, such as 
carbon nanotubes, atomic chains or semiconductor wires in scanning tunneling
spectroscopy (STS) experiments.
However, the detailed signatures can only be 
partly explained by standard Luttinger liquid theory. In particular, we find that 
the effective boundary exponent can be negative in finite wires, which 
leads to an increase of the local DOS near the edges in contrast to the
established behavior in the thermodynamic limit.
\end{abstract}


\maketitle

Interacting one-dimensional quantum wires are well studied
examples of systems in which the Fermi liquid paradigm of 
electron-like quasi-particles is known to break down.
Luttinger liquid theory predicts that strong correlations lead to the remarkable 
phenomenon of separate spin- and charge-density waves as the fundamental 
collective excitations in one dimension at low energies \cite{Voit}.
Experimental confirmation for this picture has long been
controversial, but now there is some evidence for
separately dispersing spin and charge resonances 
for quantum wires on semiconductor
hetero-structures \cite{Auslaender05,Jompol09}, quasi one-dimensional
crystals \cite{Kim06} and self-organized atomic chains \cite{Segovia99}
as a function of momentum.  
The density of states (DOS) has also been analyzed
by scanning tunneling spectroscopy (STS) in
carbon nanotubes \cite{Lee04,Venema99,Lemay01} and 
self-organized atomic chains \cite{Blumenstein11}.
This immediately invites the question,
if there are characteristic signatures from standing waves of separate 
spin and charge densities, 
that can potentially be detected in {\it locally} resolved STS experiments in finite wires.
Detailed Luttinger liquid calculations near boundaries
exist, which predict corresponding wave-like modulations in the local DOS and
a sharp reduction of the DOS near boundaries \cite{Eggert1996,Eggert00,Anfuso03,paata,Schneider10}.
However, it is far from clear if these signatures
are robust in realistic lattice systems, since even 
minimalistic models like the Hubbard chain
are not perfect Luttinger liquids.  
The two main reasons for possible discrepancies are
that first of all the assumed degeneracy of spin and charge modes in 
the low energy spectrum can never be exact and will be lifted by band curvature and
other effects.
Secondly, it is known that strong logarithmic corrections will arise
from a spin umklapp operator which is generically present in 1D electron
systems with SU(2) invariant interactions \cite{Affleck89,giamarchi,lukyanov98}.

The most relevant minimalistic lattice model is the Hubbard chain
\begin{equation}
	H = -t \sum_{\sigma,\, x=1}^{L-1}
	\left( \psi_{\sigma, x}^\dagger \psi_{\sigma, x+1}^\pdag + \hc \right) 
		+ U \sum_{x=1}^L n_{\uparrow, x} n_{\downarrow, x}
	\label{H}
,
\end{equation}
which captures the main aspects of interacting one dimensional electron systems.
This model shows signatures of spin-charge separation in numerical simulations of the  
momentum resolved DOS \cite{Benthien04}, which is the central 
quantity relevant for photoemission experiments.  In this paper we will now
analyze the {\it local} DOS as a function of position and energy which 
in turn is relevant for STS experiments.

Recently, several numerical density matrix renormalization group (DMRG) 
studies considered the local DOS
for {\it spinless} lattice fermion models in one dimension
\cite{Schneider08,Pruschke11,Jeckelmann12}, where there 
is no spin and charge separation.  Therefore, the complications mentioned above 
do not arise and the agreement with theory is close to perfect in that case.
As we will show here, the local DOS for a {\it spinful} electron system
away from half-filling is much more complex
and some key features 
of the Luttinger liquid theory are strongly renormalized.

The observable of interest is the local DOS to tunnel
an electron with spin up at a certain energy $\omega$ into the wire,
\begin{align}
	\label{LDOS}
	\rho(x, \omega) 
	&= \sum_\alpha \, \left|
	\bra{\omega_\alpha} \psi_{\uparrow,x}^\dagger \ket{0}
	\right|^2
	\,\delta(\omega - \omega_\alpha) \\
	&= \frac{1}{\pi} \Im \int_0^\infty e^{\ii \omega t} 
\ii \langle  \psi_{\uparrow,x}^\pdag(t) \psi_{\uparrow,x}^\dagger(0)  \rangle
\dd t . \label{GF}
\end{align}
Here, the fermion operator $\psi_{\uparrow,x}^\dagger$ creates a particle at position $x$ on top 
of the
groundstate $\ket{0}$ and 
the sum runs over all states $\ket{\omega_\alpha}$ 
with one additional electron.

In the following we will compare analytical calculations for the local DOS
from Luttinger liquid theory with 
simulations using the numerical 
DMRG algorithm \cite{White92,White93}. 
For the matrix elements in Eq.~(\ref{LDOS}) we 
use a multi-target DMRG with a large number of target states 
in two different particle number sectors \cite{Schneider08} and keep
track of all local fermion creation operators.

\begin{figure}
	\begin{center}
		\includegraphics[width=0.48\textwidth]{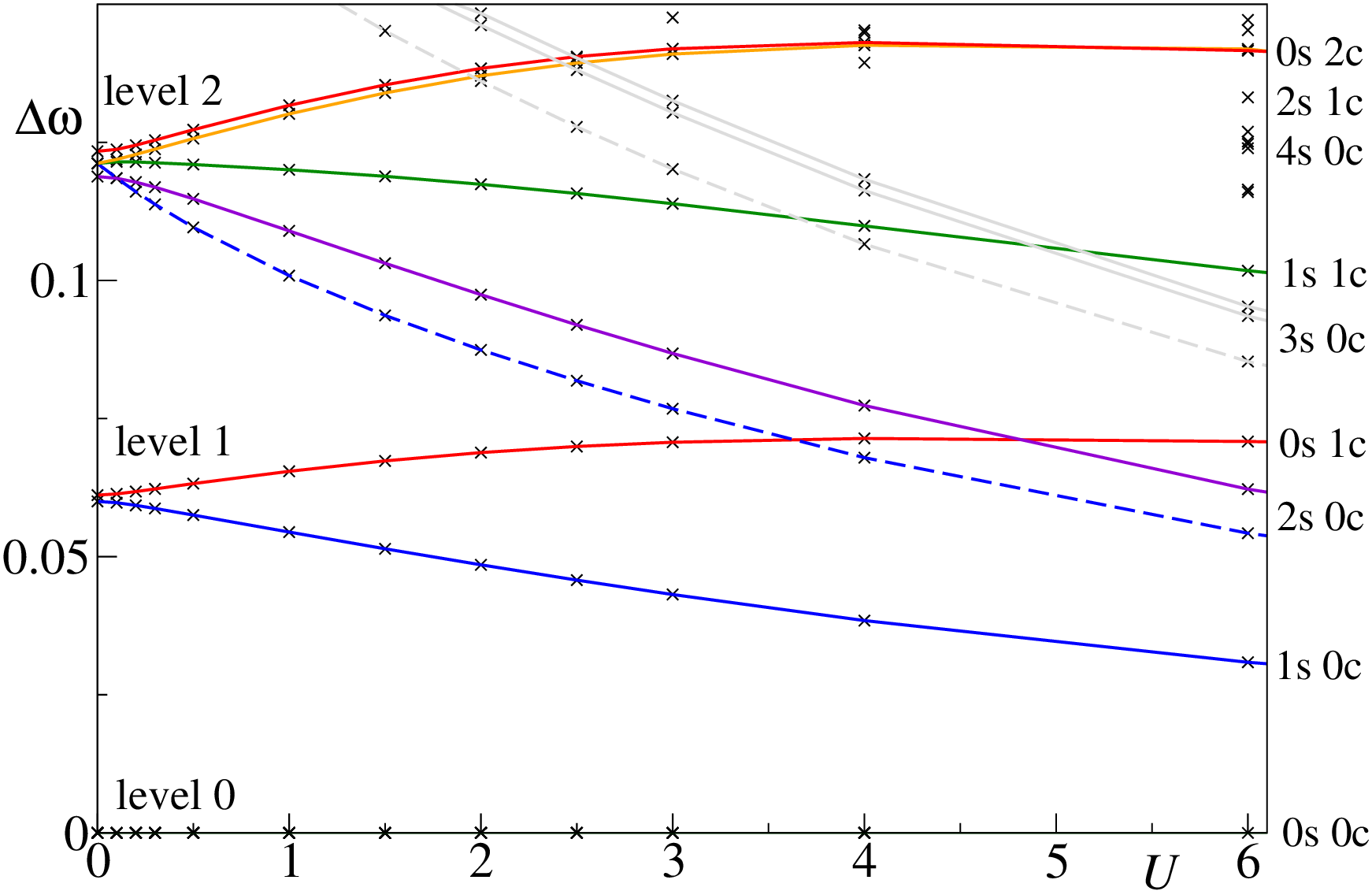}
	\end{center}
	\caption{(Color online) Energies $\Delta \omega=\omega_\alpha-\omega_0$ as a function of $U$
	for excited states with $N_\uparrow = N_\downarrow+1 = 31$ particles 
on a lattice of length $L=90$.
	The quantum numbers on the right indicate the corresponding mode $\{m_s,m_c\}$.
Dashed lines correspond to $S=\nicefrac{3}{2}$ states.   For larger $U$, excitations from level 3 (3s,0c) 
and level 4 also appear in the low energy spectrum.}
	\label{fig:excitations}
\end{figure}
In order to identify the separate spin and charge  excitations, we first 
focus on the energy spectrum $\Delta\omega=\omega_\alpha-\omega_0$ for finite wires of length $L$
as a function of interaction $U$ as shown 
in Fig.~\ref{fig:excitations} in units of $t=1$. 
The ground state $|0\rangle$ is assumed to be
a filled Fermi sea with 
total magnetization $S^z =0$, so that all particle excited states $|\omega_\alpha\rangle$
have $N_\uparrow = N_\downarrow+1$ particles and total spin z-component $S^z=\nicefrac{1}{2}$.

For $U=0$ all excitations are 
described by simple products of fermion operators 
$c_{\sigma,n}^\dagger = \sqrt{\frac{2}{L+1}}\sum_x \psi_{\sigma,x}^\dagger \sin (k_F+ 
k_{n+1})x$ where $k_n= n\frac{\pi}{L+1}$ and 
$|\omega_0\rangle = c^\dagger_{\uparrow,0}|0\rangle$ is the lowest energy
particle excitation.
Since the spectrum is approximately
linear $\Delta \omega \sim v_F (k-k_F)$, a state 
involving several fermion operators (i.e.~a multi-particle excitation)
is nearly degenerate with a single particle excitation at $U=0$ if the sum of 
excited wavenumbers is the same, which results in  
quantized fermion levels shown in Fig.~\ref{fig:excitations}. For example 
in level 1, the two states $c^\dagger_{\uparrow,1}|0\rangle$ and 
$c^\dagger_{\uparrow,0} c^\dagger_{\downarrow,0} 
c^{\phantom{\dagger}}_{\downarrow,-1}|0\rangle$ both have 
approximately the same energy $\omega_1-\omega_0 \approx v_F k_1$.
  In general there are many multi-particle states in 
each level $n$, but only one single particle state $c^\dagger_{\uparrow,n}|0\rangle$
carries all the DOS if $U=0$.

\begin{figure}
	\includegraphics[width=0.48\textwidth]{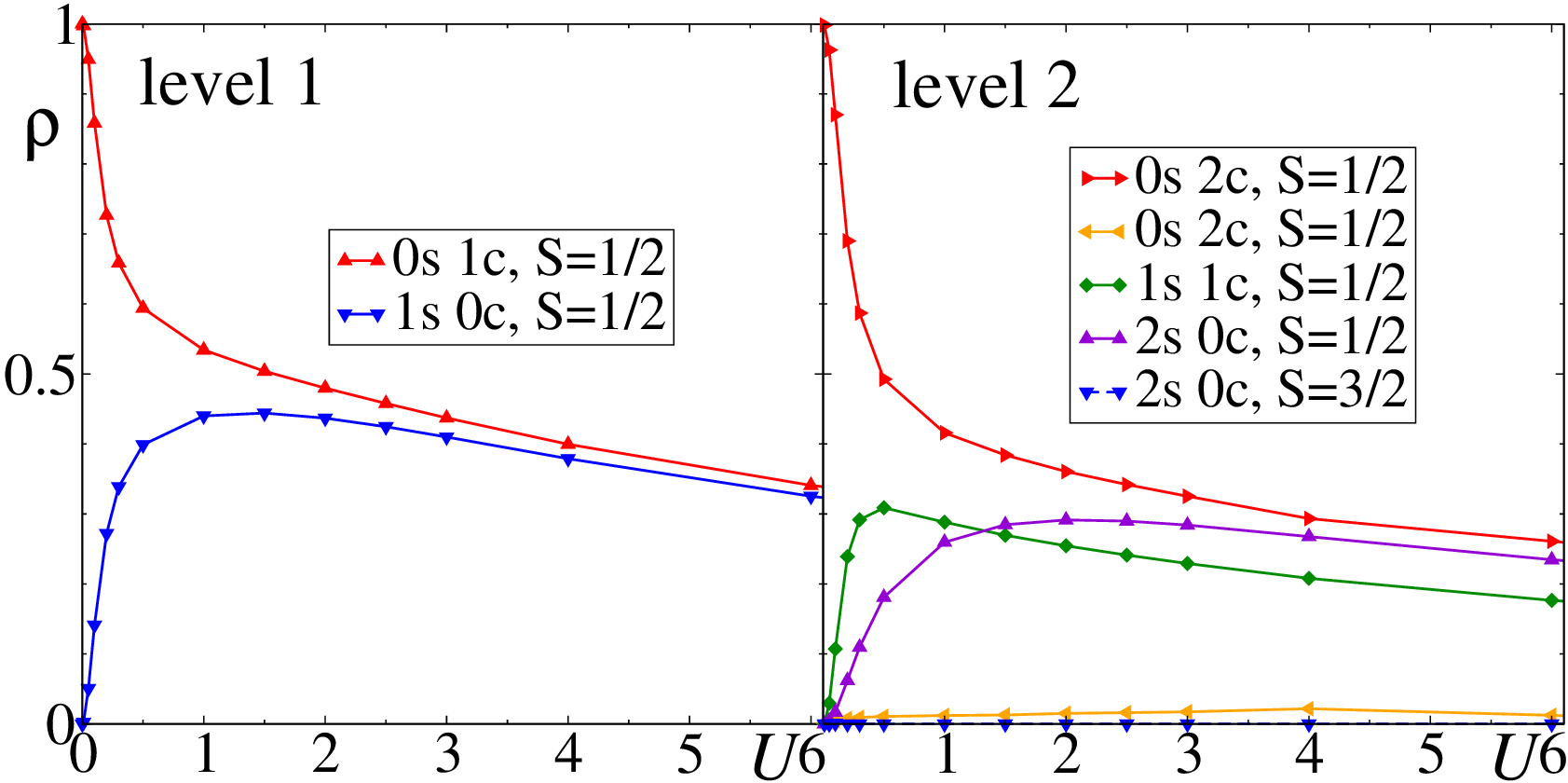}
	\caption{(Color online)
		Total DOS for the excitations of the first two levels from 
Fig.~\ref{fig:excitations}.}
	\label{fig:idos}
\end{figure}

The situation changes for finite $U$.  Now all states may potentially carry 
spectral weight in the DOS and
the near-degeneracy of the levels is lifted.  
According to the Luttinger liquid picture the states are now described by 
integer spin and charge quantum numbers $\{m_s,m_c\}$ with 
energies  $\omega_{m_s,m_c}=(m_{s}v_s+m_c v_c) \frac{\pi}{L+1} $ 
in terms of the spin and charge
velocities $v_s \leq v_c$ \cite{Voit,Schneider10}.
The spectrum in  Fig.~\ref{fig:excitations} shows a regular spin and charge spacing
with quantum numbers shown on the right.
For each spin/charge mode $\{m_s,m_c\}$ there can be several states with approximately
the same energy.  The number of states in each mode is given by the number of ways
it can be created by boson creation operators $b^\dagger_{m,\nu}$, 
i.e.~the product of the integer
partitions of $m_s$ and $m_c$.
For example, for the mode $0s\ 2c$ in Fig.~\ref{fig:excitations}, we have $m_c=2=1+1$
corresponding to the two states
created by $b_{2,c}^\dagger$ and $(b_{1,c}^\dagger)^2$, respectively, which indeed
have almost the same energy.  
On the other hand, the near degeneracies of spin modes (e.g. $2s\ 0c$)
are significantly split in Fig.~\ref{fig:excitations} 
by well-known logarithmic corrections as will be discussed below \cite{Affleck89,giamarchi,lukyanov98}.

Regarding the DOS it is now interesting to explore 
how the total spectral weight is distributed among 
the excited states.  The compact answer from Luttinger liquid theory 
is that the summed up DOS in each mode $\{m_s,m_c\}$ should 
be proportional to a 
powerlaw of the corresponding energy, 
but there are no predictions how the DOS is distributed among the states within each mode.
For the lowest energy modes the situation is still simple, since the first 
level corresponds to one spin and one charge state, which should have 
 roughly the same DOS of $\nicefrac{1}{2}$ each for small $U$ according to theory. 
However, the numerical results in Fig.~\ref{fig:idos} 
already demonstrate obvious deviations from this prediction, since the charge
mode has a much larger DOS for small $U$.  
The reason for this discrepancy comes from the non-linear band curvature, which lifts the 
degeneracy for finite $L$ even at $U=0$, so that the interaction has to 
overcome this energy splitting. 
Indeed for cases
where the two states are exactly degenerate at $U=0$ (e.g. in the thermodynamic limit)
the spin and charge states have comparable DOS even for infinitesimally small
 $U$.  As can be seen in Fig.~\ref{fig:idos} for $L=90$ the charge state dominates 
for  $U \alt 0.5$, which would imply that 
for $U\alt 10 \frac{v_s \pi}{L}$ the band curvature is dominant over 
the interaction effects.  
The next five states from the second fermion level 
in Fig.~\ref{fig:idos} show a similar crossover behavior with $U$.  
Nonetheless, features in the {\it local} DOS 
will clearly show the interaction effects even in the crossover region
 as we will see below.

In the $0s\ 2c$ and the $2s\ 0c$ modes 
there are two states each  as expected.
However, only one of the states in each mode contributes the 
overwhelming majority 
of the DOS, while the other state
would be practically invisible in an STS experiment.  
The reason that some states in a given mode
have zero DOS can sometimes be linked to exact symmetries, such as the SU(2) symmetry of 
generic Coulomb interactions. 
In particular, all particle excitations $|\omega_\alpha\rangle$ in the  
$S^z=\nicefrac{1}{2}$ sector are representatives of SU(2) multiplets.
The lowest energy state $|\omega_0\rangle$ always belongs to a 
doublet with $S=\nicefrac{1}{2}$.  
Excitations with 
charge bosons $b^\dagger_{\ell,c}$ never change the total spin, but excitations with 
spin bosons $b^\dagger_{\ell,s}$ may generate higher spin values, which 
can be calculated using the commutation rules of the non-abelian SU(2)-Kac-Moody 
algebra \cite{Affleck89}, since the spin bosons correspond to the modes 
of the SU(2) current along the z-direction as discussed in the appendix.
For example the lowest energy $S=\nicefrac{3}{2}$ state is given by the spin boson excitation
$\frac{1}{\sqrt{3}}[(b^\dagger_{1,s})^2 - b^\dagger_{2,s}] 
|\omega_0\rangle $ with $m_s=2$ which is
plotted as a dashed line in Fig.~\ref{fig:excitations}.  
Such a total spin analysis can be performed for a large number of 
excitations (see also appendix), which 
is useful since 
states with $S>\nicefrac{1}{2}$ must have zero DOS 
due to angular momentum addition rules.
For example in the 
fifth spin mode $5s\ 0c$, there are 7 states, three of which have total 
spin $S=\nicefrac{3}{2}$ and
exactly zero DOS.  Interestingly, three more states carry only very small spectral
weight, so that only one state dominates for this mode.  This indicates that 
the eigenstates remain the same in terms of their bosonic expressions even for $U\neq 0$.
For charge modes there is also exactly one state which 
carries the overwhelming weight in each mode, but the DOS of the other states 
is finite and generally increases with $U$.

For a complete analysis it is now useful to turn to the {\it local} DOS  
 $\rho_{m_s,m_c}(x)$ for each mode $\{m_s,m_c\}$.
Bosonization predicts that the 
local DOS for a given mode is composed of
a uniform and an oscillating product of spin and
charge contributions \cite{Schneider10}
\begin{eqnarray}
	 \rho_{m_s,m_c}(x)  & = & 
|c_x|^2 \left[ \rho^\uni_{s,m_s}(x)\rho^\uni_{c,m_c}(x) \right.
\nonumber \\
& & \left.- \cos(2\kF x) 
\rho^\osc_{s,m_s}(x) \rho^\osc_{c,m_c}(x) \right]
	\label{rho}
	. 
\end{eqnarray}
In what follows we will focus only on the uniform part of the local DOS 
in Eq.~(\ref{rho}), since the 
quickly oscillating $2k_F$-terms cannot be easily resolved 
in an STS experiment yet.  
The amplitudes $\rho^{\text{u}}_{\nu,m}(x)$ are slowly varying and 
can be predicted by a
simple recursive formula for spin and charge
($\nu=c,s$) 
separately \cite{Schneider10},
\begin{eqnarray}
\rho^\uni_{\nu, m}(x) & = &  \frac{1}{m} \sum_{\ell=1}^m \rho^\uni_{\nu, m-\ell}(x) \gamma^\uni_{\nu,\ell}(x),
\label{rho_recursive} \\
{\rm where~~~~ }	\gamma^\uni_{\nu,\ell}(x) &= & a_\nu + b_\nu \, \cos(2k_\ell  x). 
	\label{gammas}
\end{eqnarray}
Here we have defined spin and charge exponents $a_\nu = (1/K_\nu + K_\nu)/4$
and $b_\nu = (1/K_\nu - K_\nu)/4$ in terms of the respective
Luttinger parameters $K_\nu$ for $\nu=c,s$.
The overall prefactor 
$|c_x|^2\propto (\sin \frac{\pi x}{L+1})^{b_c+b_s}$
in Eq.~(\ref{rho}) does not depend on energy and serves as normalization so that
$\rho^{\text{u/o}}_{\nu,m=0}=1$. 
It is straight-forward to see that 
the recursive formula
results in powerlaws for the DOS in the bulk $\rho \propto \omega^{a_c+a_s-1}$
for $L\to \infty$ \cite{Schneider10,Mattsson1997}.
In addition, the formula predicts slow wavelike modulations in the local DOS 
due to the second term in Eq.~(\ref{gammas}), which also survive in the 
thermodynamic limit near the edge \cite{Eggert1996}.

\begin{figure}
	\includegraphics[width=0.48\textwidth]{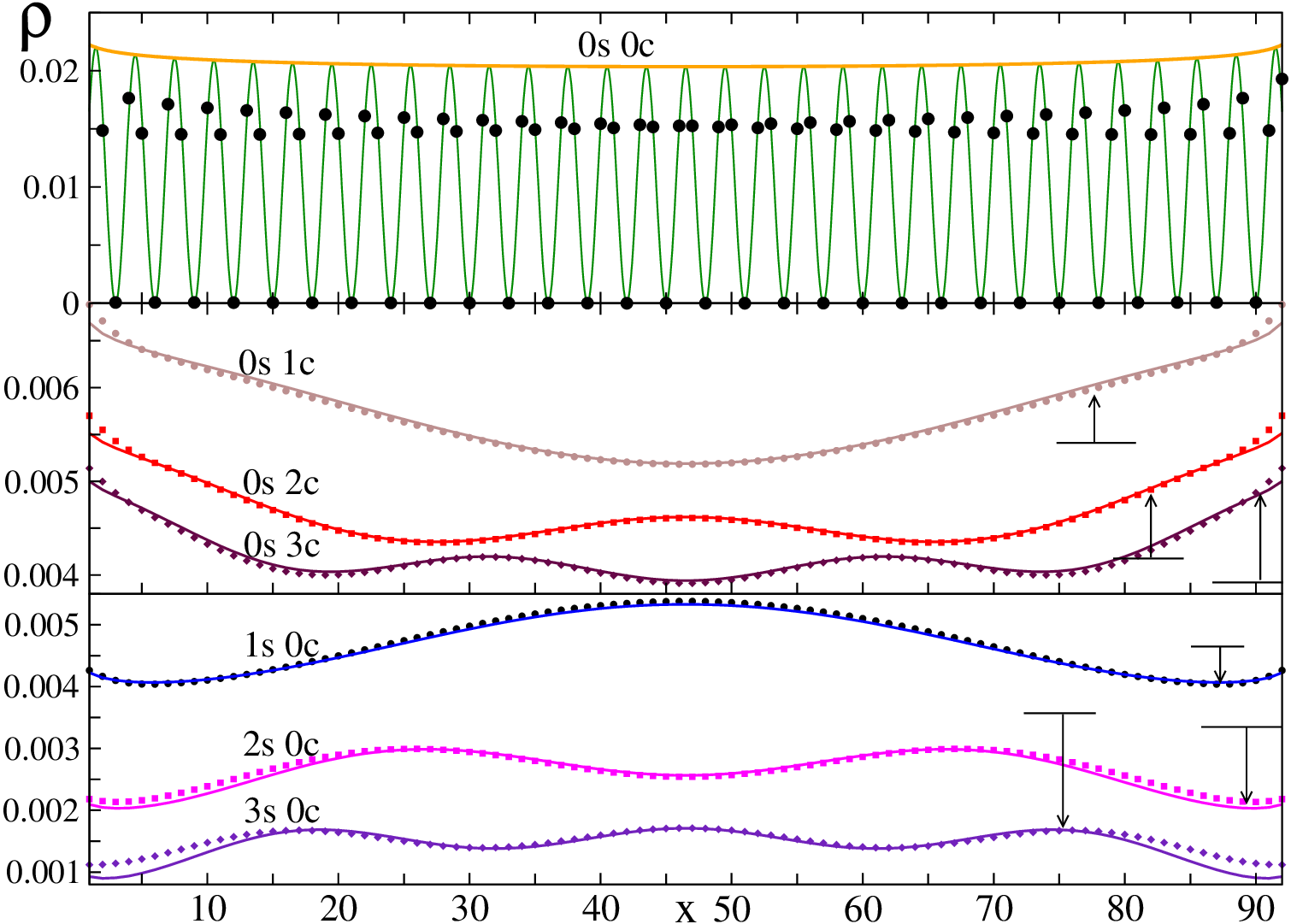}
	\caption{(Color online) The local DOS of the first few modes for $L=92$ and $U=1$.
Points are DMRG data and lines are theoretical 
predictions for $K_c=0.9081$ and $K_s=1.16$ adjusted by a shift as indicated by arrows 
(see text).
Top: Local DOS for $|\omega_0\rangle$.  The thick line 
corresponds to $2|c_x|^2$. Lower plots: Uniform part of the local DOS 
for the first few charge and spin modes. }
	\label{fig:ldos}
\end{figure}

The local DOS from the DMRG data is shown in Fig.~\ref{fig:ldos} 
for the first few modes at $\nicefrac{1}{3}$-filling.  For the lowest excitation $|\omega_0\rangle$
the oscillating and uniform parts
are the same and given by the prefactor $|c(x)|^2$. 
Already at first sight it is surprising to see that the local DOS for all modes
increases slightly near the boundary, 
while previous works have predicted it
to decrease according to the boundary exponent\cite{Kane1992,Eggert1996,Mattsson1997}.  
Indeed it must 
be emphasized that the local DOS does {\it not} fit the theoretical prediction.
All curves should in principle be fit free, up to one overall normalization, since
the local DOS of all levels follows from Eqs.~(\ref{rho})-(\ref{gammas}), where the
Luttinger parameters $K_c(U)$ and $K_s=1$ are known from the thermodynamic 
Bethe ansatz. 
However, in  Fig.~\ref{fig:ldos} two important adjustments have been made:
First the theoretical curves were shifted
down for the charge modes and up for the spin modes in order
to fit the numerical data (indicated by arrows).
This adjustment was already observed in the crossover of the total DOS in Fig.~\ref{fig:idos}
due to the competition of energy scales (band curvature vs.~interaction)
as argued above.
Secondly, we find that the spin Luttinger liquid parameter must be
chosen considerably larger than unity $K_s\approx 1.16$ for all spin modes in
order to fit the numerical data 
corresponding to {\it attractive} behavior in the spin modes.
This is especially surprising since the charge Luttinger parameter from 
Bethe ansatz
$K_\charge = 0.9081$ agrees perfectly with the data without any finite size adjustments.
There are no other adjustable parameters in the fits of Fig.~\ref{fig:ldos}, except for one
overall normalization constant.
In return, the results show that knowing the local DOS from (numerical) experiments,
it is possible to extract the effective parameters $K_s$ and $K_c$ 
from the modulations of only the first spin and charge excited states 
according to Eqs.~(\ref{rho})-(\ref{gammas}).
  The oscillating parts $\rho^\osc_\nu$ can be analyzed
analogously and give the same results (not shown).

\begin{figure}
	\includegraphics[width=0.48\textwidth]{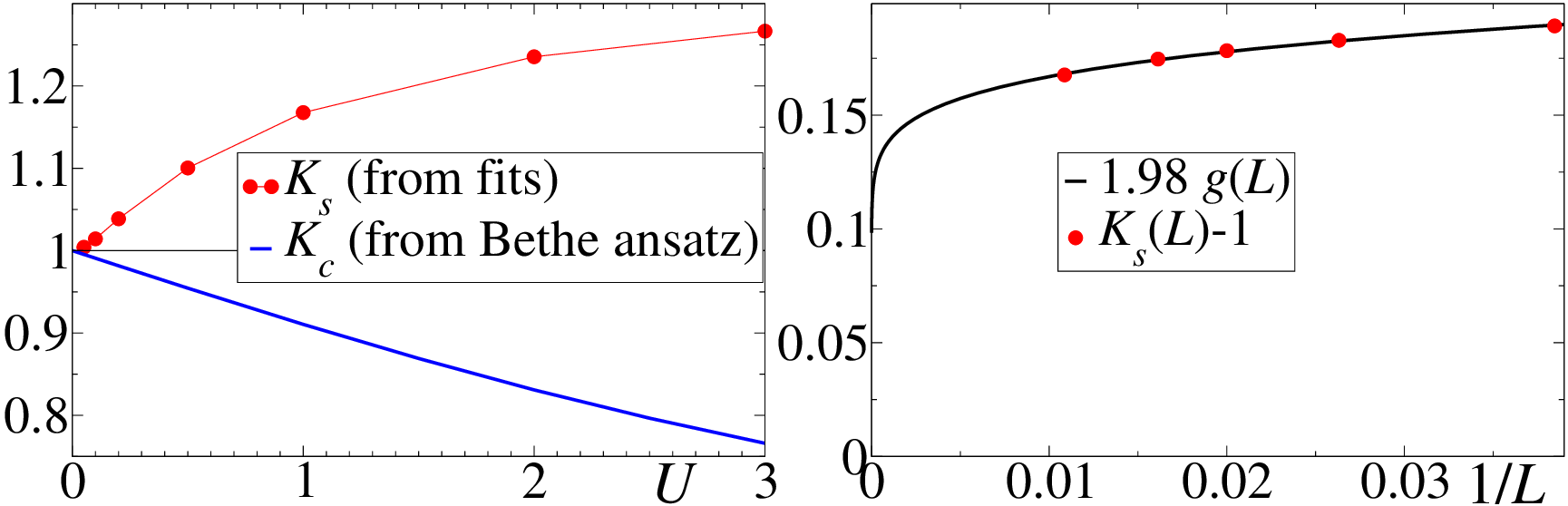}
	\caption{(Color online) Left:
Luttinger parameters $K_\charge$ (line) from Bethe ansatz and $K_\spin$ 
(points) from the fits to $\rho^\uni_{s, m=1}(x)$ as a function of $U$ for $L=92$. 
	Right: Renormalization of $K_\spin$ (points) at $U=1$ with the system size $L$
compared to Eq.~\eqref{Lukyanov}. }
	\label{fig:Kscaling}
\end{figure}

In Fig.~\ref{fig:Kscaling} the behavior of the Luttinger parameters from 
the corresponding fits to the local DOS
is shown as function of interaction and length at $\nicefrac{1}{3}$-filling.
The charge parameter from Bethe ansatz $K_c$ always agrees
very well with the data without any additional adjustments.
However, 
the observed spin parameter $K_s$ is considerably larger than
$K_s=1$. 
Non-abelian bosonization predicts $K_s=1$ for 
any SU(2) invariant model, 
but at the same time it is known that a marginal irrelevant operator
causes corrections to the anomalous dimension which only vanish logarithmically slowly 
with $1/\ln L$ in the thermodynamic limit \cite{Affleck89,giamarchi}.  
In abelian bosonization such a correction 
can indeed effectively be modeled by a renormalizing spin 
Luttinger parameter \cite{Affleck89,giamarchi,lukyanov98,edgesusc,sirker2008}
\begin{equation}
K_s-1\propto  g, \text{ with }
	g^{-1} + \frac{1}{2} \ln(g) = \ln(L/L_0)
	\label{Lukyanov}
\end{equation}
where $L_0$ 
is non-universal and depends on the model and the quantity of interest. 
As shown in Fig.~\ref{fig:Kscaling} such a renormalization description is 
indeed consistent with our data for $K_s$. 
The parameter $K_s$ increases with $U$
at a given length $L$, but decreases slowly as the length is increased.
The parameter $K_s$ appears to be the same for all spin modes 
at a given $U$ and $L$, 
i.e.~independent of energy $\omega$.  
The renormalization of $K_s$ is very slow, so that exponentially large systems
are required to observe the thermodynamic limit $K_s\to 1$.  The fit parameters for 
$K_s \approx 1+ 1.98 g$
with $\ln L_0 \approx -6$ are outside the range what would normally be expected 
for a spin chain model \cite{lukyanov98}.  
Therefore, the particular form of the observed corrections remains a puzzle.

Nonetheless, the results of the logarithmic corrections have 
interesting consequences.  In particular, the corrections are so large, that 
the boundary exponent $\alpha_{B}= (1/K_s+1/K_c)/2-1$ may become negative
if $K_s+K_c < 2 K_s K_c$, i.e.~$K_s-1 \agt 1-K_c$ 
to lowest order in the correction, which is indeed the case 
for small $U \alt 2t$ at $L=92$.
This results in an {\it increase} of the local DOS 
for small energies $\omega$ and small distances $x$ near the boundary,
which is described by a weak powerlaw 
divergence $\rho \propto x^{b_s+b_c} \omega^{\alpha_B}$.
Such a negative boundary exponent would also explain the recently observed anomalies
in the boundary behavior for small interactions 
in functional renormalization group studies \cite{Schuricht12}.

In conclusion, we have analyzed the local DOS of the Hubbard model in the low energy regime.
Individual states can be classified by separate spin- and charge quantum numbers. 
We observe that typically only one eigenstate 
contributes the overwhelming majority of the local DOS in each
spin/charge mode.  The spin and charge Luttinger parameters $K_s$ and $K_c$
can be extracted from the modulations in the local DOS of individual excited states. 
While the charge
parameter $K_c$ agrees well with the Bethe ansatz, the 
spin Luttinger liquid parameter is attractive $K_s>1$
due to large finite size corrections, which 
can only be neglected for exponentially long chain lengths.
In fact, the corrections to $K_s$ are
unexpectedly strong and may even lead to negative boundary exponents for moderate 
interactions $U\alt 2t$. The common assumption that it is possible to generically 
use $K_s=1$ due to SU(2) invariance is certainly not justified for the local DOS.
In particular, for finite wires on conducting substrates the interactions may be
reduced by screening, so that 
the charge Luttinger liquid parameter $K_c$ may be close to one, while the spin
Luttinger liquid parameters $K_s$ can already be significantly increased, which
leads to a negative boundary exponent.  This would have quite dramatic consequences, since
the DOS near the boundary determines the tunneling between connected wires and 
the renormalization of the conductivity through impurities \cite{Kane1992}, 
which will show an {increase} at low temperatures in this scenario.

\begin{acknowledgments}
We are thankful for useful discussions with A.~Struck and M.~Bortz.
This work was supported by 
the DFG and the State of Rheinland-Pfalz via
the SFB/Transregio 49 and the MAINZ graduate school of excellence.
\end{acknowledgments}

\appendix*
\section{Appendix: Non-abelian Bosonization}

\label{sec:bos:nonabelian}
In order to determine the total spin of an excitation 
it is useful to use non-abelian bosonization in the spin channel \cite{Affleck89}.
In this case the excitations are created by the modes of SU(2) currents $J^a_m$
with $a=x,y,z$ obeying the Kac-Moody algebra
\begin{equation}
		[{J^a_m},{J^b_n}]
		= \ii \varepsilon^{abc} J^c_{m+n} + \tfrac{1}{2}m \delta^{a,b} \delta_{m,-n}
\label{kacmoody}
\end{equation}
The ground state is 
characterized by $J^a_m|\omega_0\rangle =0, \ \forall\, m<0$.
The total spin operator is given 
in terms of the $m=0$ currents
\begin{equation}
S^2 = \vec{J}_0 \cdot \vec{J}_0 = 2J_0^+J_0^- + J_0^z + (J_0^z)^2
\end{equation}
where $J^{\pm} = J^x \pm \ii J^y$.   
The current modes in the z-direction are related to the abelian spin bosons above by 
$J^z_m = \sqrt{\frac{m}{2}} b^\dagger_{m,s} $ 
and $J^z_{-m} = \sqrt{\frac{m}{2}} b_{m,s} $ for $m>0$.
It is therefore straight-forward to consider the total spin of any
bosonic spin and charge excitation by using the Kac-Moody commutation relations.
The charge bosons commute with the total spin operator $S^2$.   
Spin excitations are created by products of spin creation operators 
$b^\dagger_{m,s}$ acting on $|\omega_0\rangle$.  The corresponding 
normalized states can be
labelled by the set of which bosons were created
$\{m_1,m_2,m_3,....\}\rangle$, 
e.g.  $|\{3,1,1\}\rangle = \frac{1}{\sqrt{2}} b^\dagger_{3,s} (b^\dagger_{1,s})^2 |\omega_0\rangle$.
Therefore, the matrix elements of $\langle \{m_1,m_2,m_3,...\}| S^2| \{m_1',m_2',m_3',...\}\rangle$ 
between any two such excitations can be evaluated 
uniquely by the Kac Moody algebra (\ref{kacmoody}).
The $J^z_0$ operators commute with all excitations and the
ground state is characterized by $J^z_0 |\omega_0\rangle = S^z|\omega_0\rangle= \frac{1}{2}|\omega_0\rangle$
in our case.  For the $J^\pm$ operators we use the Kac-Moody relation in 
Eq.~(\ref{kacmoody}) with the help of computer algebra in order to successively commute them to the right until the action on the ground state is known.  This results in a
non-diagonal matrix for $S^2$ for each spin mode separately, which can be 
brought into diagonal form.  The resulting eigenstates and eigenvalues are as 
follows: \begin{widetext}
{	\begin{equation*}
		\begin{array}[t]{c|l}
			\mathbf{S=\nicefrac{1}{2}} & {\ket{\omega_0}} \\
			\hline
			S=\nicefrac{1}{2} & \ket{\{1\}} \\
			\hline
			S=\nicefrac{1}{2} & \sqrt{\frac{2}{3}} \ket{\{2\}} + \sqrt{\frac{1}{3}} \ket{\{1,1\}} \\
			S=\nicefrac{3}{2} & -\sqrt{\frac{1}{3}} \ket{\{2\}} + \sqrt{\frac{2}{3}} \ket{\{1,1\}} \\
			\hline
			S=\nicefrac{1}{2} & \sqrt{\frac{1}{3}} \ket{\{3\}} + \sqrt{\frac{2}{3}} \ket{\{1,1,1\}} \\
			S=\nicefrac{1}{2} & \sqrt{\frac{2}{9}} \ket{\{3\}} + \sqrt{\frac{6}{9}} \ket{\{2,1\}} - \sqrt{\frac{1}{9}} \ket{\{1,1,1\}} \\
			S=\nicefrac{3}{2} & -\sqrt{\frac{4}{9}} \ket{\{3\}} + \sqrt{\frac{3}{9}} \ket{\{2,1\}} + \sqrt{\frac{2}{9}} \ket{\{1,1,1\}}\\
			\hline
			S=\nicefrac{1}{2} & -\sqrt{\frac{1}{3}} \ket{\{3,1\}} + \sqrt{\frac{2}{3}} \ket{\{1,1,1,1\}} \\
			S=\nicefrac{1}{2} & \sqrt{\frac{12}{27}} \ket{\{4\}} + \sqrt{\frac{2}{27}} \ket{\{3,1\}} + \sqrt{\frac{12}{27}} \ket{\{2,1,1\}} + \sqrt{\frac{1}{27}} \ket{\{1,1,1,1\}} \\
			S=\nicefrac{1}{2} & \sqrt{\frac{3}{54}} \ket{\{4\}} + \sqrt{\frac{8}{54}} \ket{\{3,1\}} + \sqrt{\frac{27}{54}} \ket{\{2,2\}} - \sqrt{\frac{12}{54}} \ket{\{2,1,1\}} + \sqrt{\frac{4}{54}} \ket{\{1,1,1,1\}} \\
			S=\nicefrac{3}{2} & -\sqrt{\frac{1}{3}} \ket{\{4\}} + \sqrt{\frac{1}{3}} \ket{\{2,2\}} + \sqrt{\frac{1}{3}} \ket{\{2,1,1\}} \\
			S=\nicefrac{3}{2} & -\sqrt{\frac{1}{6}} \ket{\{4\}} + \sqrt{\frac{4}{9}} \ket{\{3,1\}} - \sqrt{\frac{1}{6}} \ket{\{2,2\}} + \sqrt{\frac{2}{9}} \ket{\{1,1,1,1\}} \\
		\end{array}
	\end{equation*}}
for the case that the state $|\omega_0\rangle$ has 
total spin of $S=\nicefrac{1}{2}$.  \\~\\
It is also possible to consider 
a $|\omega_0\rangle$ state with $S=0$.  In that case the eigenstates are given by:
	\renewcommand{\arraystretch}{1.2}
	\setlength{\tabcolsep}{10pt}
	\setlength{\arraycolsep}{10pt}
	\begin{equation*}
		\begin{array}[t]{c|l}
			\mathbf{S=0} & {\ket{\omega_0}} \\
			\hline
			S=1 & \ket{\{1\}} \\
			\hline
			S=0 & \ket{\{1,1\}} \\
			S=1 & \ket{\{2\}} \\
			\hline
			S=0 & \ket{\{2,1\}} \\
			S=1 & \ket{\{3\}} \\
			S=1 & \ket{\{1,1,1\}} \\
			\hline
			S=0 & \sqrt{\frac{2}{3}} \ket{\{3,1\}} + \sqrt{\frac{1}{3}} \ket{\{1,1,1,1\}} \\
			S=0 & \sqrt{\frac{1}{9}} \ket{\{3,1\}} + \sqrt{\frac{2}{3}} \ket{\{2,2\}} - \sqrt{\frac{2}{9}} \ket{\{1,1,1,1\}} \\
			S=1 & \ket{\{4\}} \\
			S=1 & \ket{\{2,1,1\}} \\
			S=2 & -\sqrt{\frac{2}{9}} \ket{\{3,1\}} + \sqrt{\frac{1}{3}} \ket{\{2,2\}} + \sqrt{\frac{4}{9}} \ket{\{1,1,1,1\}} \\
			\hline
			S=0 & \sqrt{\frac{4}{7}} \ket{\{4,1\}} + \sqrt{\frac{3}{7}} \ket{\{2,1,1,1\}} \\
			S=0 & \sqrt{\frac{6}{63}} \ket{\{4,1\}} + \sqrt{\frac{49}{63}} \ket{\{3,2\}} - \sqrt{\frac{8}{63}} \ket{\{2,1,1,1\}} \\
			S=1 & \ket{\{5\}} \\
			S=1 & \ket{\{3,1,1\}} \\
			S=1 & \ket{\{2,2,1\}} \\
			S=1 & \ket{\{1,1,1,1,1\}} \\
			S=2 & -\sqrt{\frac{1}{3}} \ket{\{4,1\}} + \sqrt{\frac{2}{9}} \ket{\{3,2\}} + \sqrt{\frac{4}{9}} \ket{\{2,1,1,1\}} \\
		\end{array}
	\end{equation*}
\end{widetext}
\end{document}